\documentclass[letterpaper, 10 pt, conference]{ieeeconf}  

\IEEEoverridecommandlockouts                              

\usepackage{blindtext}

\usepackage{verbatim,multirow,hhline}
\usepackage[usenames,dvipsnames]{xcolor}

  \usepackage[pdftex]{graphicx}
\usepackage[cmex10]{amsmath}
\usepackage{subfigure}
\usepackage{tikz}
\usepackage[utf8]{inputenc}
\usepackage{amsfonts}
\usepackage{amssymb}
\usepackage{multirow}
\usetikzlibrary{shapes,arrows, fit}
\usetikzlibrary{positioning}

\tikzstyle{response} = [cloud, circle, draw, fill=blue!10, 
    text width=5em, text centered,  minimum height=3em]

\tikzstyle{level1} = [cloud, circle, draw, fill=white!10, 
    text width=4em, text centered,  minimum height=2em] 
    
\tikzstyle{desire} = [cloud, circle, draw, fill=red!10, 
    text width=4em, text centered,  minimum height=2em] 

\tikzstyle{level2} = [cloud, circle, draw, fill=red!10, 
    text width=4em, text centered,  minimum height=2em]

\tikzstyle{arrow} = [ultra thick,->,>=stealth]
\tikzstyle{a} = [rectangle, draw, minimum height=17em, minimum width=15em]

\title{\LARGE \bf
Decoding Complex Imagery Hand Gestures*
}

\author{Seyed Sadegh Mohseni Salehi$^{2}$, Mohammad~Moghadamfalahi, Fernando Quivira \\ Alexander Piers,
Hooman Nezamfar, and Deniz~Erdogmus$^{1}$
\thanks{*Our work is supported by NSF (IIS-1149570, CNS-1544895), NIDLRR (90RE5017-02-01), and NIH (R01DC009834). Relevant code and data will be disseminated via Northeastern University Digital Repository Service, in collection NEU/COE/ECE/CSL (permalink: http://hdl.handle.net/2047/D20199232).}
\thanks{$^{1}$ {Electrical and Computer Engineering Department, Northeastern University, Boston}}%
\thanks{$^{2}$ {\tt\small ssalehi@ece.neu.edu}}
}

\begin{document}

\maketitle
\thispagestyle{empty}
\pagestyle{empty}

\begin{abstract}

Brain computer interfaces (BCIs) offer individuals suffering from major disabilities an alternative method to interact with their environment. Sensorimotor rhythm (SMRs) based BCIs can successfully perform control tasks; however, the traditional SMR paradigms intuitively disconnect the control and real task, making them non-ideal for complex control scenarios. In this study we design a new, intuitively connected motor imagery (MI) paradigm using hierarchical common spatial patterns (HCSP) and context information to effectively predict intended hand grasps from electroencephalogram (EEG) data. Experiments with 5 participants yielded an aggregate classification accuracy--intended grasp prediction probability--of 64.5\% for 8 different hand gestures, more than 5 times the chance level.

\end{abstract}

\section{INTRODUCTION}

Brain Computer Interfaces have shown promises in providing individuals with alternative interaction methods via brain activities. 
Major electroencephalogram (EEG) based BCI systems can be categorized into three different categories based on the type of brain activity they utilize, event-related potential (ERP) based, steady state visually evoked potential (SSVEP) based, and motor imagery (MI) based.

ERP and SSVEP are more popular in BCI systems where the short response time and the availability of several options are of concern. Controlling a wheelchair~\cite{nezamfar2013brain} and typing with characters~\cite{nezamfar2016flashtype} are two examples that rely on the above features, respectively. In control and navigation applications SSVEP-based BCI systems are widely used due to their fast response. 
SSVEP systems are fast, but they require the visual stimulation and vision capabilities~\cite{nezamfar2015stimuli} not suitable for individuals with severely impaired vision.

MI, the process of imagining a physical movement without execution, can be categorized as imagining moving different limbs (simple MI) and imagining different movements of the same limb (compound MI)\cite{yi2013eeg}. MI tasks desynchronize the stable, resting alpha/mu ($7-12$ Hz) and beta ($12-30$ Hz) rhythms, that can be measured through EEG potentials on the scalp\cite{yi2013eeg,pfurtscheller1999event}. The spatial distribution of brain activity measured by EEG signals in response to these imaginations depend on the complexity of the MI task; simple MI (imagination of left hand vs. right leg) produces activity in different spatial locations while compound MI (open palm vs. closed fist) produces activities in similar spatial cortical areas~\cite{vuvckovic2012two}. The high spatial correlation of compound MI coupled with the volume conduction effect, the spatial distortion of EEG signals on the scalp caused by travelling through tissue and fluid, make different MI tasks difficult to distinguish.

One particular application of BCIs is the control of prosthetics or exoskeletons for individuals with motor control loss.  From the BCI perspective, a brain controlled hand requires rapidly (real-time) and accurately classified MI that maps directly to the motion of the target limb. Simple MI is not sufficiently intuitive, and compound MIs are difficult to differentiate. Hence classifying EEG potentials suitable to control a robotic limb is a challenge.

Several studies have classified gestures with moderate success, but these gestures are not sufficiently intuitive to be used in a BCI prosthetic. These studies investigated classifying simple MI by simply distinguishing imagination of right and left hand movements with $80\%-93\%$ accuracy\cite{vuvckovic2012two,pfurtscheller1997eeg,hamedi2014neural}. Others have completed binary classifications on different same-limb movements, such as distinguishing between wrist flexion, extension, supination, pronation, and finger movement\cite{vuvckovic2012two,vuckovic2008delta,mohamed2011single} resulting in accuracies between $58-82\%$. However, the higher accuracies are only achieved when wrist extension has been one of the binary options; without wrist extension classification accuracies tend to run closer to the lower side of the reported range. Binary classification does not provide enough degrees of freedom to control a prosthetic hand, and the imaginary movements used for classification are not necessarily intuitive hand motions. A recent study~\cite{edelman2016eeg} classified four hand states (wrist flexion, extension, supination, and pronation) using source imaging analysis~\cite{edelman2014discriminating} to improve signals' separability. Even with this combination, the imagined  wrist motions do not map directly or intuitively enough to natural hand movements.
In this paper, we propose classification of four grasps on each hand totaling to 8 options. The grasps mimic four of the required grasps to complete everyday tasks. The goal is to improve classification accuracy and time to control a prosthetic hand, utilizing the natural choice of grasps.

\section{Method}

\subsection{Experiment design}
Classifier parameters are estimated using data collected during a supervised calibration session. The classifier is meant to discriminate EEG signals corresponding to imagination of four gestures shown in Fig.~1.a. Prior to data collection, participants get familiarized with gestures through executing each gesture two times, for 5 seconds. During the calibration session, participants are presented with the \emph{target} gesture for 2 seconds during which they are asked to focus on that particular gesture. Then, upon the presentation of the word \emph{start} on the display participants imagine the target gesture for 5 seconds (Fig.~1.b). Here, each MI task for a gesture is called a \emph{trial}. 
Inter trial interval in the calibration task is arbitrary and is decided by the user based on their required time to rest. Each participant performs 20 trials of each gesture, summing to a total of 160 MI trials.

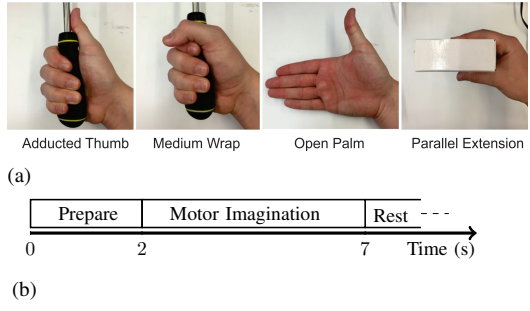
\begin{figure}
    \centering
     \begin{subfigure}[]
         \centering
         \includegraphics[width=0.8\columnwidth]{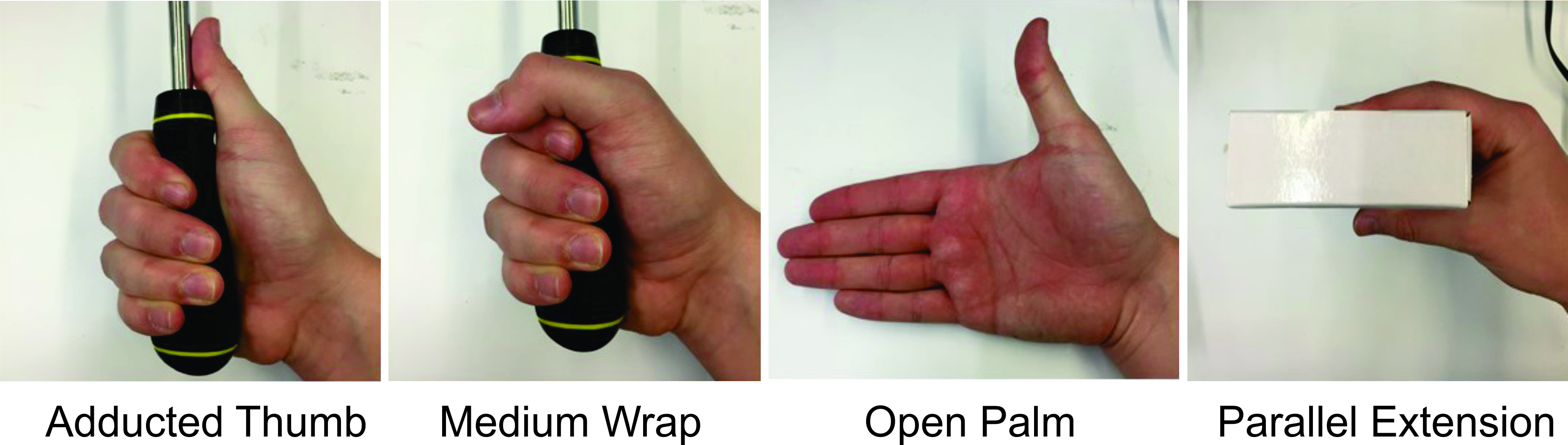}
     \end{subfigure}
          \vfill
            \begin{subfigure}[]
             \centering
             \resizebox{.8\columnwidth}{!}{
             \begin{tikzpicture}
                 \draw[ultra thick,->] (0,0) -- (8,0);
                 \draw[dashed] (7,0.35) -- (7.5,0.350);
                 \foreach \x in {0,2}
                 \draw (\x cm,1pt) -- (\x cm,-1pt) node[anchor=north]{$\x$};
                 \draw (6 cm,1pt) -- (6 cm,-1pt) node[anchor=north]{$7$};
                 \draw[thick] (0,0.1) rectangle (2,0.6);
                 \node[text width=2cm] at (1.5,0.32) {Prepare};
                 \node[text width=4cm] at (4.5,0.32) {Motor Imagination};
                 \node[text width=1cm] at (6.65,0.32) {Rest};
                 \draw[thick] (2,0.1) rectangle (6,0.6);
                 \draw[thick] (6,0.1) -- (7,0.1);
                 \draw[thick] (6,0.6) -- (7,0.6);
                 \node[text width=2cm] at (7.75,-0.3){Time (s)};
             \end{tikzpicture}
             }
             \caption{ a) gestures for each hand. b) The experimental paradigm}
        \end{subfigure}
    \label{fig:4gestures}
\end{figure}

\subsection{EEG feature extraction}
Prior to feature extraction, EEG signals are filtered with a $3-30$~Hz FIR band-pass filter to retain the alpha/mu and beta frequency bands, as they show the most activity in response to imagination and execution of hand gestures~\cite{pfurtscheller1997eeg}. Assuming the band-passed EEG signal follows a Gaussian distribution within the time window, and source activity constellations between different class pairs are independent, we can design a spatial filter (spatial transform) that maximizes the signal's variance for one class and minimizes for the other. We used the common spatial pattern~(CSP) algorithm~\cite{ramoser2000optimal} to calculate the spatial transforms. CSP algorithm is used to find the transformation matrix between class pairs.
CSP finds the best coordinates between binary classes.
In the multiclass case, we propose a Hierarchical Common Spatial Pattern~(HCSP) to utilize a series of binary transformations found by CSP.

\begin{figure}
   \centering
   \resizebox{.8\columnwidth}{!}{
\begin{tikzpicture}[sibling distance=9em,
  branch/.style = {shape=rectangle, rounded corners,
    draw, align=center,
    top color=white, bottom color=blue!20}]]
  \node[branch] {Decision}
    child { node[branch] {Left Hand}
        child { node[branch] {Fingers Extension}
            child { node[branch] {Thumb Abduction}}
            child { node[branch] {Thumb Adduction}}}
        child { node[branch] {Fingers Flexion}}}
    child { node[branch] {Right Hand}};

\node[text width=1cm,font=\bf] at (2,-2.1) {$\ddots$};
\node[text width=1cm,font=\bf] at (0.4,-3.6) {$\ddots$};
  \node[text width=1cm] at (-1.5,-0.75) {$L^1$:};
    \node[text width=1cm] at (-3.5,-2.25) {$L^2$:};
    \node[text width=1cm] at (-5.5,-3.75) {$L^3$:};
\end{tikzpicture}
}
        \caption{Hierarchical classification model}%
        \vspace{-.5cm}
        \label{fig:Hierarchy}%
\end{figure}
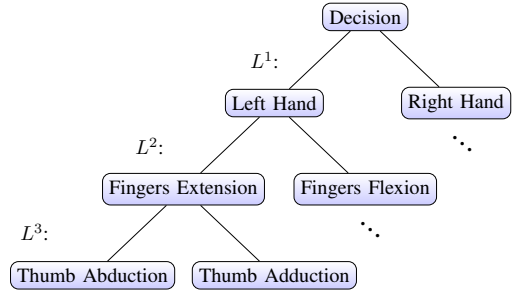

 In HCSP, classes with common properties are placed in one category. For example, first level contains two categories, motions on the right hand and motions on the left hand. Similarly, extension or flexion of fingers and abduction or adduction of thumbs are considered different categories in the second and the third level respectively. Fig~\ref{fig:Hierarchy} outlines the HCSP for classification of different hand gestures. The goal of each classifier is to classify the categories not classes. By cascading the category classifiers, all 8 classes can be detected.

 The normalized spatial covariance matrix is estimated as
\begin{equation}
C_j = \frac{1}{N_j}\sum_{i=1}^{N_j}\frac{E_i^{(j)}{E_i^{(j)}}'}{trace(E_i^{(j)}{E_i^{(j)}}')}
\end{equation}
where $E_i^{(j)}$ is an $m\times t$ matrix representing filtered windowed EEG evidence corresponding to the $i^{th}$ trial  of the $j^{th}$ class. Additionally, $m$ is the number of channels, $t$ is the number of samples in a time window, $'$ is the transpose operator, $trace(.)$ is the sum of the diagonal elements of the matrix, and $N_j$ is the number of trials for the $j^{th}$ class.

In order to find the spatial transformation matrix at each level,$V_L$, the optimization problem in equation~\ref{eqe:CSP} should be solved.
\begin{equation}
    V_L^* = \underset{V_L}{\operatorname{argmax}}\frac{V_L^T \sigma_{L_{l=-1}} V_L}{{V_L}' (\sigma_{L_{l=-1}}+\sigma_{L_{l=+1}}) V_L}
    \label{eqe:CSP}
\end{equation}
where $\sigma_{L_{l=-1}}$ and $\sigma_{L_{l=+1}}$ are the average covariance matrices - where $\pm1$ represent one of the categories - in level $L$ and are calculated by
\begin{equation}
    \sigma_{L_{l}} = \frac{\sum_{j\in category({L_{l}})}N_j C_j}{\sum_{j\in category({L_{l}})}N_j}
    \label{eqe:sigma}
\end{equation}
$\sum_{j\in category({L_{l}})}N_j$ is the number of trials for each category at each level.

Solving the optimization problem represented in equation~\ref{eqe:CSP} is identical to solving generalized eigenvalue problem 
\begin{equation}
    V_{L}'\sigma_{L_{l=-1}}V_L = D \wedge V_{L}'(\sigma_{L_{l=-1}}+\sigma_{L_{l=+1}})V_L = I
    \label{eqe:GEP}
\end{equation}
where $D$ is a diagonal matrix.

 According to each spatial filter $V_L$, computed for level $L$, the projected data $P_{L_l}$ corresponding to the $i^{th}$ trial of the $j^{th}$ class, $E_i^{(j)}$, can be calculated as
\begin{equation}
    P_{L_l} = {V_L}' E^{(j\in category(L_l))}
    \label{eqe:projected data}
\end{equation}
In equation~\ref{eqe:GEP} the $k$ smallest and largest eigenvalues in $D$ correspond to $k$ leftmost/rightmost columns in spatial filter $V$ respectively. These values yield to the smallest variance in category $L_{l=-1}$ and simultaneously largest variance in category $L_{l=+1}$ and vice versa. To extract the features, the first and the last $k$ rows of $P_{L_l}$ are considered. 

Using $P_{L_l}^{(K)}$ ($K=1 \hdots 2k)$ which maximizes the difference of variances between two categories, $f_{L_l}^{(K)}$, features for each trial of the corresponding category at each level are calculated as
\begin{equation}
    f_{L_l}^{(K)} = log \left (\frac{var(P_{L_l}^{(K)})}{\sum_{q=1}^{2k}var(P_{L_l}^{(q)})}\right )
    \label{eqe:feature}
\end{equation}
where, $i$ is the trial index in each the category. 
\subsection{EEG Likelihood Probability Extraction}
\label{sec:method_classification}
For proper estimation of the classifier performance a leave-one-out cross validation approach is used. At each level, the corresponding feature vector, $f_{L_{l_i}}^{(K)}$, is calculated for each trial. The classifier is trained in two steps. First, Fisher LDA algorithm is used to extract fisher scores by
\begin{equation}
F_{L_{l_i}}^{(K)} = \left (w_L^{(K)}\right )' f_{L_{l_i}}^{(K)}
\end{equation}
where
\begin{equation}
    w_L^{(K)} \propto \left (\Sigma_{L_{l={-1}}}^{(K)} + \Sigma_{L_{l={+1}}}^{(K)}\right )^{-1} \left (\mu_{L_{l={+1}}}^{(K)}-\mu_{L_{l={-1}}}^{(K)}\right )
\end{equation}
and $\Sigma_{L_{l={-1}}}^{(K)}, \mu_{L_{l={-1}}}^{(K)}$ and $\Sigma_{L_{l={+1}}}^{(K)}, \mu_{L_{l={+1}}}^{(K)}$ are the category's covariance and mean.

Then, the likelihood probability densities are calculated for the fisher scores in each category $P(F_{L_{l}}^{(K)}|L=l)$.
Multi-variate Gaussian model is considered at each level for density estimation due to the nature of EEG signals. 
Classification accuracy is defined as the probability of the correct answer.\par

Different integer window lengths from 1 to 5 seconds are considered for feature extraction. 
\subsection{Graphical model}
\label{sec:graphical_model}
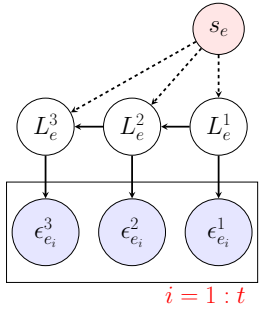
\begin{figure}%
    \centering
        \resizebox{.4\columnwidth}{!}{
\begin{tikzpicture}[thick,scale=0.9, every node/.style={transform shape}]

\tikzstyle{a} = [rectangle, draw, minimum height=10em, minimum width=25em]
 \node [level1] (hand) {\Huge $L_e^1$};
 \node [level1, left = 1cm of hand] (fingers) {\Huge $L_e^2$};
 \node [level1, left = 1cm of fingers] (thumb) {\Huge $L_e^3$};
 \node [desire, above = 1.5cm of hand] (Decision) {\Huge $s_e$};
\node [response, below = 1.5cm of  hand] (EEGHand) {\Huge $\epsilon_{e_i}^1$};
\node [response, below = 1.5cm of  fingers] (EEGFingers) {\Huge $\epsilon_{e_i}^2$};
\node [response, below = 1.5cm of  thumb] (EEGThumb) {\Huge $\epsilon_{e_i}^3$};

\draw [arrow] (hand) -- (fingers) ;
\draw [arrow,dashed] (Decision) -- (hand);
\draw [arrow,dashed] (Decision) -- (fingers);
\draw [arrow,dashed] (Decision) -- (thumb);
\draw [arrow] (fingers) -- (thumb);

\draw [arrow] (hand) -- (EEGHand);
\draw [arrow] (fingers) -- (EEGFingers);
\draw [arrow] (thumb) -- (EEGThumb);
\node[a,fit =(EEGHand) (EEGFingers) (EEGThumb) ,label={[label distance=0, text=red]300:\Huge $i=1:t$}] (container) {};
\end{tikzpicture}
}
        \caption{System graphical model in epoch $e$. The final state $s_e$ is composed of the state in each level $L_{e}^{M}$ and thus deterministically related (dashed lines). States are calculated from the observed random variables and extracted EEG features of each level $\epsilon_{e_i}^M$. States and random variables are probabilistically related (solid lines). }%
        \vspace{-.5cm}
        \label{fig:GraphicalModel}%
    \end{figure}

 The graphical model in Fig~\ref{fig:GraphicalModel} is proposed to probabilistically merge the category classifiers and employ the inter-level information to calculate the posterior probability of each gesture.
This model represents the generative model of the collected EEG data in epoch $e$ and the time required for the model to make a decision ($1-5$ s). 
The goal of this graphical model is to estimate the next state $s_e$ with incorporation of inter-level prior information.
The inter-level information is the prior probability of a gesture at a level given the state at the higher level. 
This information is gathered based on the combination of gestures used over time for different tasks.
For instance, the probability of the thumb being closed would be increased if closed fingers is detected at the higher level.
Specifically in this graphical model:$s_e$ is the decision between all classes at epoch $e$, $L_e^j$ represents the decision from the level j classifier, and $\epsilon_{e_i}^j$ is the extracted CSP feature from EEG evidence at time $i$ for decision $L_e^j$.

Using the graphical model above, the posterior probability of each state given the observed random variables ($\{\epsilon_{e_i}^1\}_{i=1}^{t},\{\epsilon_{e_i}^2\}_{i=1}^{t},\{\epsilon_{e_i}^3\}_{i=1}^{t}$) can be calculated to construct the PMF over all gestures, required to make a decision.  The posterior probabilities are calculated as follows:
\begin{equation}
\begin{split}
    P(s_e&\ |\{\epsilon_{e_i}^1\}_{i=1}^{t},\{\epsilon_{e_i}^2\}_{i=1}^{t},\{\epsilon_{e_i}^3\}_{i=1}^{t}) \\
    &\ = P(L_e^1,L_e^2,L_e^3|\{\epsilon_{e_i}^1\}_{i=1}^{t},\{\epsilon_{e_i}^2\}_{i=1}^{t},\{\epsilon_{e_i}^3\}_{i=1}^{t}) \\
    &\ \propto P(\{\epsilon_{e_i}^1\}_{i=1}^{t},\{\epsilon_{e_i}^2\}_{i=1}^{t},\{\epsilon_{e_i}^3\}_{i=1}^{t}|L_e^1,L_e^2,L_e^3)\\  
    &\ \times P(L_e^1,L_e^2,L_e^3) \\
    &\ \propto P(\{\epsilon_{e_i}^1\}_{i=1}^{t}|L_e^1)P(\{\epsilon_{e_i}^2\}_{i=1}^{t}|L_e^2)P(\{\epsilon_{e_i}^3\}_{i=1}^{t}|L_e^3) \\ 
    &\ P(L_e^3|L_e^2)P(L_e^2|L_e^1)P(L_e^1)
\end{split}
\end{equation}

In this set of equations: $P(\{\epsilon_{e_i}^M\}_{i=1}^{t}|L_e^M)$, represents EEG observation likelihood for a particular gesture at level $M$, estimated using the calibration data; $P(L_e^M|L_e^{M-1})$, is the probability of each command at level $M$ given the inter-level information; and $P(L_e^1)$, represents the prior probability of each hand being target.
\subsection{Decision Criteria}\label{sysDec}
One decision is made in every epoch. Decisions are made using a Maximum-a-Posteriori (MAP) estimate.
	\begin{equation}
		\hat{s}_e=\arg\max_{s_e}{P(s_e|\{\epsilon_{e_i}^1\}_{i=1}^{t},\{\epsilon_{e_i}^2\}_{i=1}^{t},\{\epsilon_{e_i}^3\}_{i=1}^{t})}
		\label{eq:decision}
	\end{equation}
At each time interval $t$, if the posterior probability of $\hat{s}_e$ is greater than the system \emph{Confidence Threshold}, it will be chosen as the final decision of the epoch. 
If after $N$ time interval (here $N=5$) the posterior probability is not greater than the system \emph{Confidence Threshold}, the maximum posterior probability of $\hat{s}_e$ across all time intervals will be chosen as the final decision of the epoch.
Analysis and results
\section{Analysis and Results}
Five healthy participants (3 males and 2 females) in the range of 20-30 years old consented and participated in a data collection session following an approved protocol by Northeastern University's IRB office. Participants were not under the influence of any chemicals, such as caffeine. 

Before starting the experiment, different gestures were described to the participants; participants executed these gestures to get more familiar with the motions to be imagined during the imagination task. Participants were asked to imagine the entire progress of the gestures, not only the final position. Between each trial (2 seconds preparation, 5 second gesture imagination), participants rested until they felt comfortable to continue. On average, the experiment consisting of 160 trials, 4 different gestures on each hand took less than 45 minuets. 

\begin{figure}
    \centering
    \includegraphics[width=0.9\columnwidth]{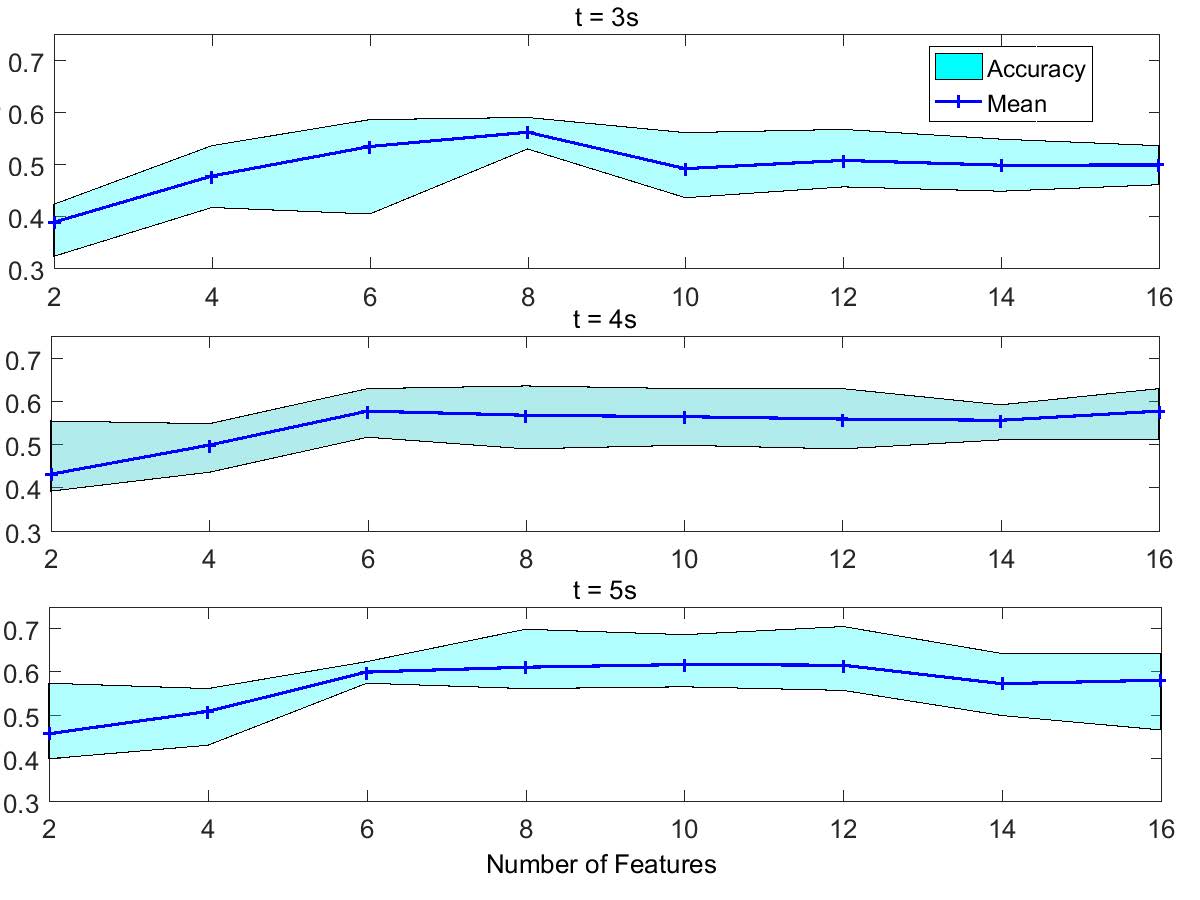}
    \caption{Average system accuracy using different number of features (x-axis) and different time-window lengths of EEG evidence.}
    \label{fig:accuracy8}
\end{figure}

\begin{figure}
    \centering
    \includegraphics[width=0.9\columnwidth]{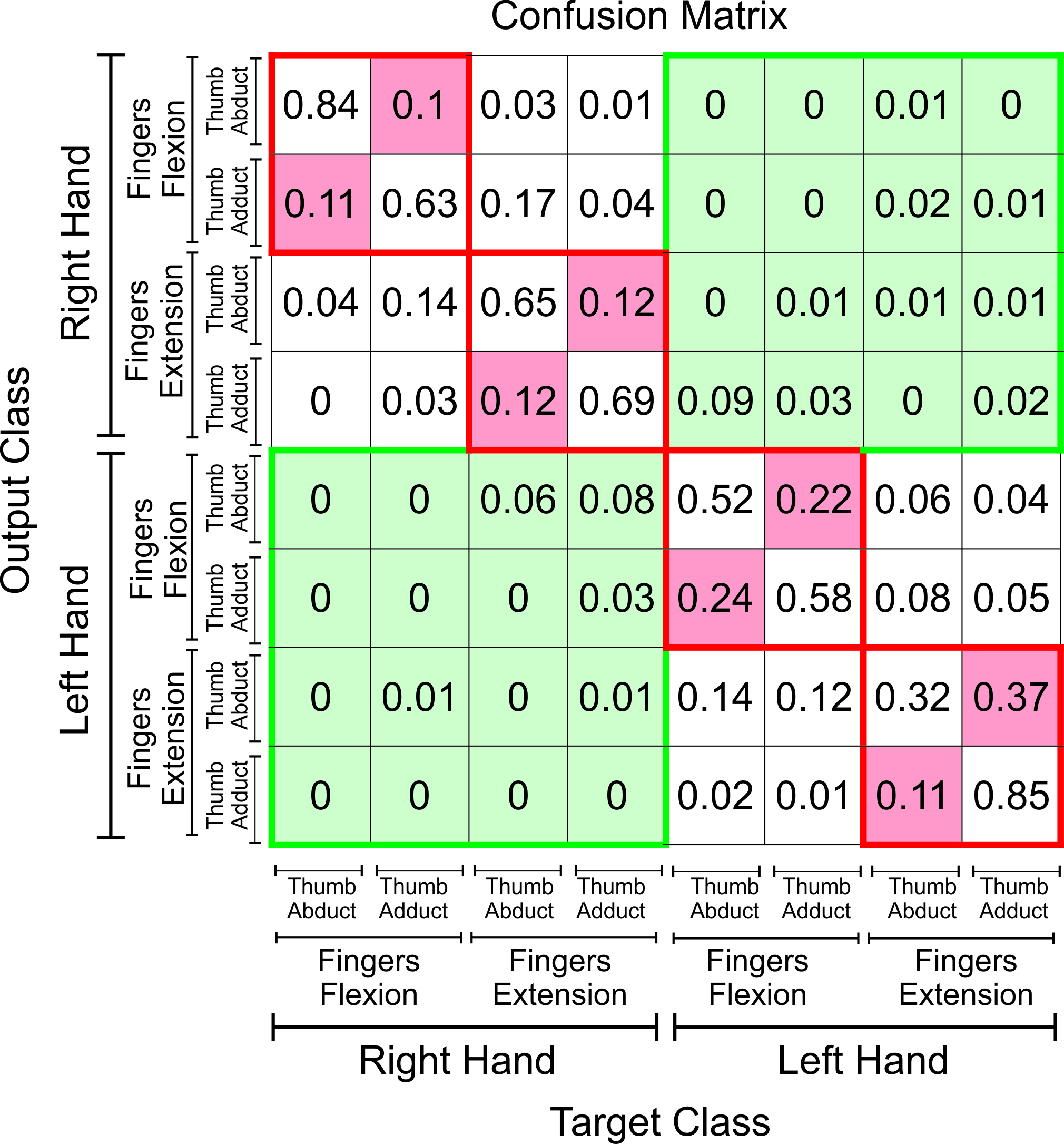}
    \caption{Average confusion matrix for 5 participants with using $K=6$ features and the window length of $t=5$ seconds}
    \label{table:8 classes}
\end{figure}

All possible number of features ($K$) and window lengths ($t\geqslant3s$) combinations were evaluated offline to find the parameter values leading to the highest classification accuracy.
Figure~\ref{fig:accuracy8} shows the total system accuracy as a function of the window length and number of features.
As expected, more EEG evidence (larger $t$) results in higher accuracy.
The average maximum accuracy of 64.5\% (71\% the highest) is achieved using $K=6$ and the window length of $t=5$ seconds using uniform inter-level prior probability. 
The average confusion matrix is shown in figure~\ref{table:8 classes}.
The confusion matrix shows the system accurately classifies between left and right hand (level 1), hence the two approximately zero blocks on the top-right and bottom-left of the matrix (Green squares).
The highest confusion is between classes that differ in thumb position (level 3), as shown by the highlighted red blocks.
The confusion matrix represented in figure~\ref{table:8 classes} shows gestures with differences in details are not as separable as the more general levels like left and right hand. 

\section{DISCUSSION}
With the goal of utilizing intuitive and natural gestures towards building motor imagery based classifiers.
The natural gestures are more complex they reflect on the same region and require more sophisticated classifiers as opposed to the tasks performed or imagined on the similar limbs on the different sides.
We proposed a hierarchical model combining binary classifiers on different levels taking advantage of the similarities in different gestures. 
Binary CSP is used to extract features at each level.
The results from each classifier is probabilistically merged, and inter-level probabilities are employed to achieve a multiclass gesture classifier. In a study with 5 healthy participants our method resulted in an average accuracy of $64.5\%$ among 8 complex hand gestures, more than 5 times the chance level.

\addtolength{\textheight}{-12cm}   



\bibliographystyle{IEEEtran}
\bibliography{refs}

\end{document}